\documentclass[10pt,a4paper,twoside]{article}
\usepackage{epsfig}
\usepackage{baltlat6}
\usepackage{array}
\usepackage{here}
\begin{document}
\ \
\vspace{0.5mm}
\setcounter{page}{277}
\vspace{8mm}

\titlehead{Baltic Astronomy, vol.\,20, 000--000, 2011}

\titleb{OFF-AXIS VARIABILITY OF AGNS: A NEW PARADIGM FOR BROAD-LINE- AND CONTINUUM-EMITTING REGIONS}

\begin{authorl}
\authorb{C. Martin Gaskell}{}
\end{authorl}

\begin{addressl}
\addressb{}{Centro de Astrof\'isica de Valpara\'iso y Departamento de F\'isica y Astronom\'ia,  Universidad de
Valpara\'iso, Av.~Gran Breta\~na 1111, Valpara\'iso, Chile; \\martin.gaskell@uv.cl}
\end{addressl}

\submitb{Received: 2011 June 10; accepted: 2011 July 26}

\begin{summary} The general picture of how thermal AGNs work has become clearer in
recent years but major observational puzzles threaten to undermine this picture. These puzzles include AGNs
with extremely asymmetric emission line profiles, inconsistent
multi-wavelength variability, rapid apparent changes in the sizes of emitting regions
and in the direction of gas flow, a curious insensitivity of gas
in some narrow velocity ranges to changes in the ionizing continuum, and
differing dependencies of polarization on gas velocity. It is proposed that all
these puzzles can readily be explained by off-axis variability.
\end{summary}


\resthead{OFF-AXIS VARIABILITY OF ACTIVE GALACTIC NUCLEI}
{C.~M. Gaskell}

\sectionb{1}{INTRODUCTION}

AGNs can be divided into ``thermal AGNs'' and ``non-thermal AGNs'' (see Antonucci 2011) depending on whether the dominant continuum emission is thermal or non-thermal.  Thermal AGNs have high accretion rates while non-thermal AGNs have accretion rates many orders of magnitude below the Eddington limit.  I will argue here that we now have a reasonably secure picture of what the inner regions of a thermal AGN are like, but that there are many hard-to-explain observations.  I will discuss how continuum variability of thermal AGNs {\em has} to be non-axisymmetric on short timescales and then outline how this off-axis variability readily explains the hard-to-explain observations.

\sectionb{2}{THE STANDARD MODEL OF THERMAL AGNS}

Our current paradigm of thermal AGNs (and probably non-thermal AGNs as well) is that the energy ultimately comes from accretion of matter onto supermassive black holes. The accreting material forms an accretion disc (Lynden-Bell 1969).  As is reviewed in Gaskell (2008), the emission of thermal AGNs is dominated energetically by thermal emission of the so-called ``big blue bump'' (BBB).  A small amount of the BBB radiation is Compton up-scattered to give hard X-ray emission, and some of it is absorbed by dust grains and re-radiated in the infra-red.  The BBB radiation is also reprocessed in the broad emission lines that are almost certainly present in all thermal AGNs.

I believe that you do not understanding something unless you know what it looks like.
Although it was (or should have been) clear from Curtis (1918) and Jennison \& Das Gupta (1953) that AGNs have axial symmetry, models of AGNs were mostly spherically symmetric until the late 1970s.  A paradigm shift took place between 1978 and 1980 when Blandford \& Rees (1978) proposed that BL Lac objects were AGNs seen with their beams aimed at us, and Keel (1980) discovered that Seyfert 1 galaxies (lower luminosity AGNs where we can see the broad-line region, BLR) are preferentially face on.  These discoveries led to what are called ``unified theories'' of AGNs.  In these theories AGNs are fundamentally similar but look different depending on their orientation because of either relativistic beaming (``beaming unification'') and/or obscuration by a dusty torus (``obscuration unification'').  These theories are thoroughly reviewed by Antonucci (1993, 2011).  It has long been recognized from the temperature of the dust that the radius of the torus is somewhat greater than the radius of the BLR.  This is confirmed by reverberation mapping measurements (Clavel, Wamstecker, \& Glass 1989; for more recent results see Suganuma et al.\@ 2006 and Gaskell et al.\@  2007).  Gaskell, Klimek, \& Nazarova (2007; GKN) argue that the BLR is simply an inner extension of the dusty torus and show that the BLR is flattened, highly stratified in ionization, and must have a similar covering factor to the torus.  For more details see Gaskell (2009).  A cartoon of the BLR and torus is shown in Fig.~1.


\begin{figure}[!tH]
\vbox{
\centerline{\psfig{figure=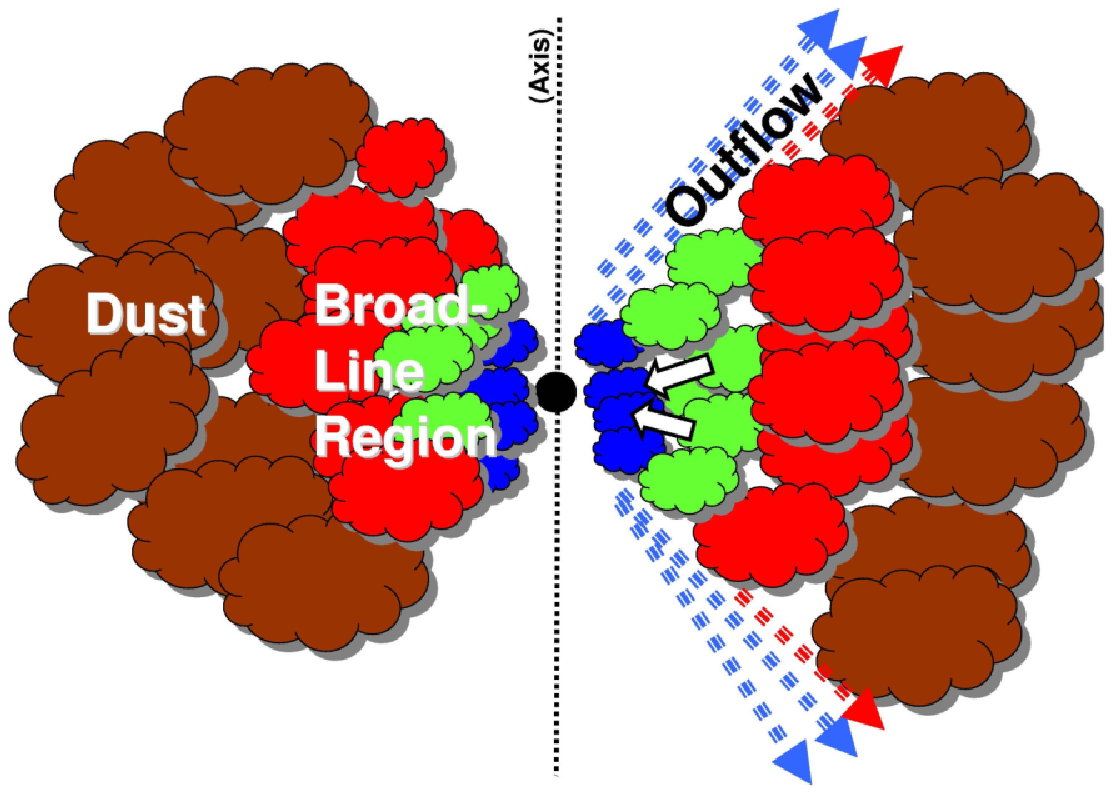,width=80mm,angle=0,clip=}}
\vspace{1mm}
\captionb{1}
{\footnotesize A schematic view of the BLR and torus of an AGN in a plane through the axis.  The figure is approximately to scale (except that the black hole is shown too large!). Figure from Gaskell, Goosmann, \& Klimek 2009.}
}
\end{figure}

The kinematics have to be consistent with this structure.  As proposed by Osterbrock (1978), the dominant motion in AGNs is rotation and then there is a substantial vertical component of velocity (what Osterbrock called ``turbulence'').  Gaskell \& Goosmann (2008) argue that there is also a significant amount of inflow.  This geometry and kinematics not only explains double-peaked ``disc-like'' broad-line profiles, but also, contrary to what is widely believed, the model also readily explains centrally-peaked ``logarithmic'' profiles (see Gaskell 2010).  A relatively small change of viewing angle is all that is needed.  This is illustrated in Fig.~2.


\begin{figure}
\vbox{
\centerline{\psfig{figure=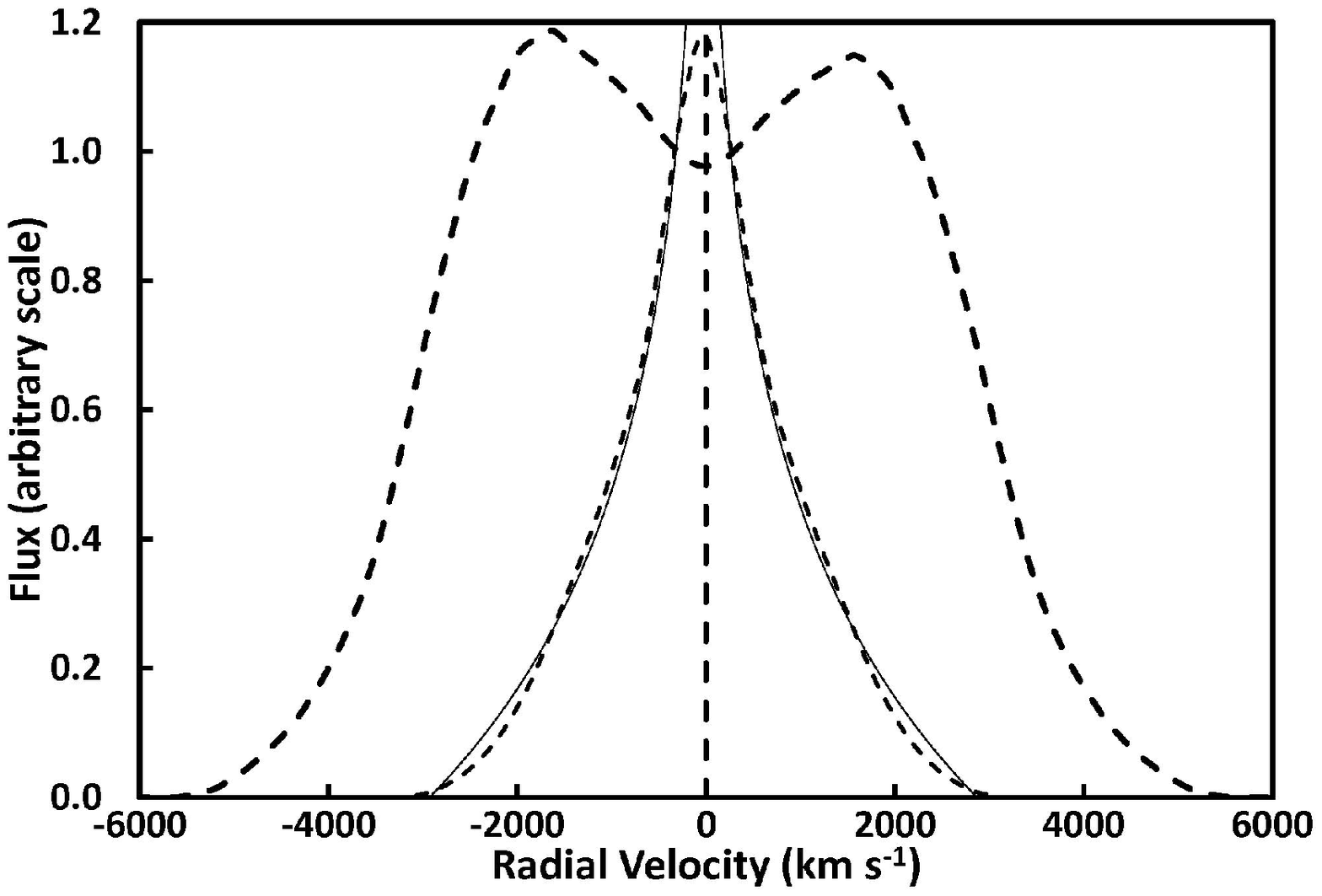,width=80mm,angle=0,clip=}}
\vspace{1mm}
\captionb{2}{\footnotesize The dot-dashed line shows the Balmer line profile that would arise from the GKN model if it were centrally illuminated and viewed from 30$^\circ$ off axis.  The profile with short dashes is the same model viewed from face-on. The thin line superimposed on this is a logarithmic profile.
Figure from Gaskell (2010).}
}
\vspace{3mm}
\label{}
\end{figure}

\sectionb{3}{CHALLENGES FOR THE CURRENT PARADIGM}

While the GKN model is consistent with unified models of AGNs and explains broad-line strengths and profiles, there are many puzzling observations which have no ready explanation in the current axisymmetric paradigm (Gaskell 2010).  These include the following:

\begin{enumerate}
\item A large fraction of AGNs have asymmetric broad line profiles and some have {\em extremely} asymmetric  profiles with the emission line peaks displaced from the rest of the host galaxy by thousands of km~s$^{-1}$ (Gaskell 1983; Gaskell 1996; Eracleous \& Halpern 2003).
\item Multi-wavelength monitoring shows that different spectral regions often vary independently and in a way that is inconsistent with reprocessing of higher-energy radiation (see Gaskell 2006, 2008 and references therein).
\item In reverberation mapping the discrepancies between the line responses and continuum variability exceed the observational errors and different continuum variability events give different time delays (Maoz 1994).
\item Velocity-resolved reverberation mapping can show discrepant kinematic signatures (Denney et al.\@ 2009) and there can be rapid apparent changes in the apparent direction of gas motion on timescales that are much too short to be real (Kollatschny \& Dietrich 1996).
\item Root-mean-square spectra can show that the most variable part of a line profile can have a surprisingly narrow range of velocity (see Fig.~1f of Denney et al.\@ 2010, for example).
\item Broad line gas in narrow velocity ranges frequently shows puzzling insensitivities to changes in the ionizing continuum (Sergeev et al.\@ 2001). This is shown in Fig.~3. It can be seen that each observing season there are narrow velocity regions showing significantly weaker correlations or even anti-correlations with the continuum flux.


\begin{figure}
\vbox{
\centerline{\psfig{figure=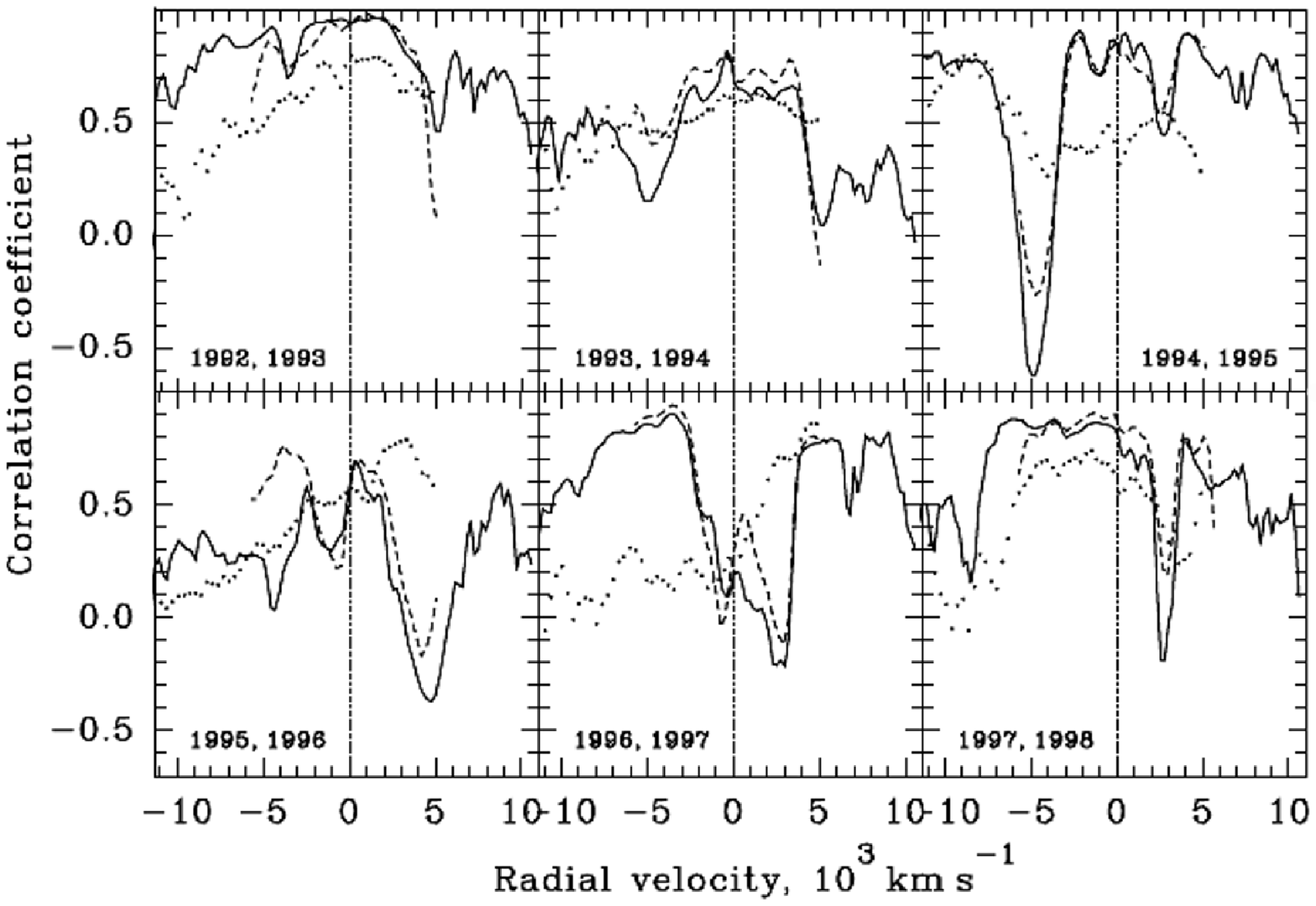,width=110mm,angle=0,clip=}}
\vspace{1mm}
\captionb{3}{\footnotesize The correlation of broad-line flux with continuum flux as a function of velocity for NGC~4151 in six separate observing seasons. The solid line shows correlation coefficients for H$\alpha$, the dashed line for H$\beta$, and the dots for He\,II $\lambda$4686. Adapted from Sergeev et al.\@ (2001).}
}
\label{}
\end{figure}

\item The polarization changes systematically across broad lines, but the changes imply different kinematics in different objects (see Axon et al.\@ 2008).
\end{enumerate}

\sectionb{4}{OFF-AXIS VARIABILITY}

Contrary to the assumptions of the axisymmetric paradigm, continuum variability {\em has} to be non-axisymmetric (Gaskell 2008).  The reason for this is easy to understand.  First let us consider an axisymmetric accretion flow.  The idealized radial temperature gradient produced in an accretion flow when the only heating comes from the potential energy liberated at a given radius has long been understood and the resultant spectrum calculated (Pringle \& Rees 1972; Shakura \& Sunyaev 1973; PRSS).  In reality there will be additional heating at a given radius because of the flow of heat from the hotter inner regions.  This flattens the temperature gradient.  The left panel of Fig.~4 illustrates the temperature gradients resulting from pure local energy generation (the case considered by PRSS) and central illumination only; the left panel shows the resulting spectra.  Note the strong sensitivity of the spectral shape to the temperature gradient.  This means that the temperature gradient is well constrained by the observed spectral shape.


\begin{figure}[htbp]
  \begin{minipage}[b]{65 mm}
    \psfig{figure=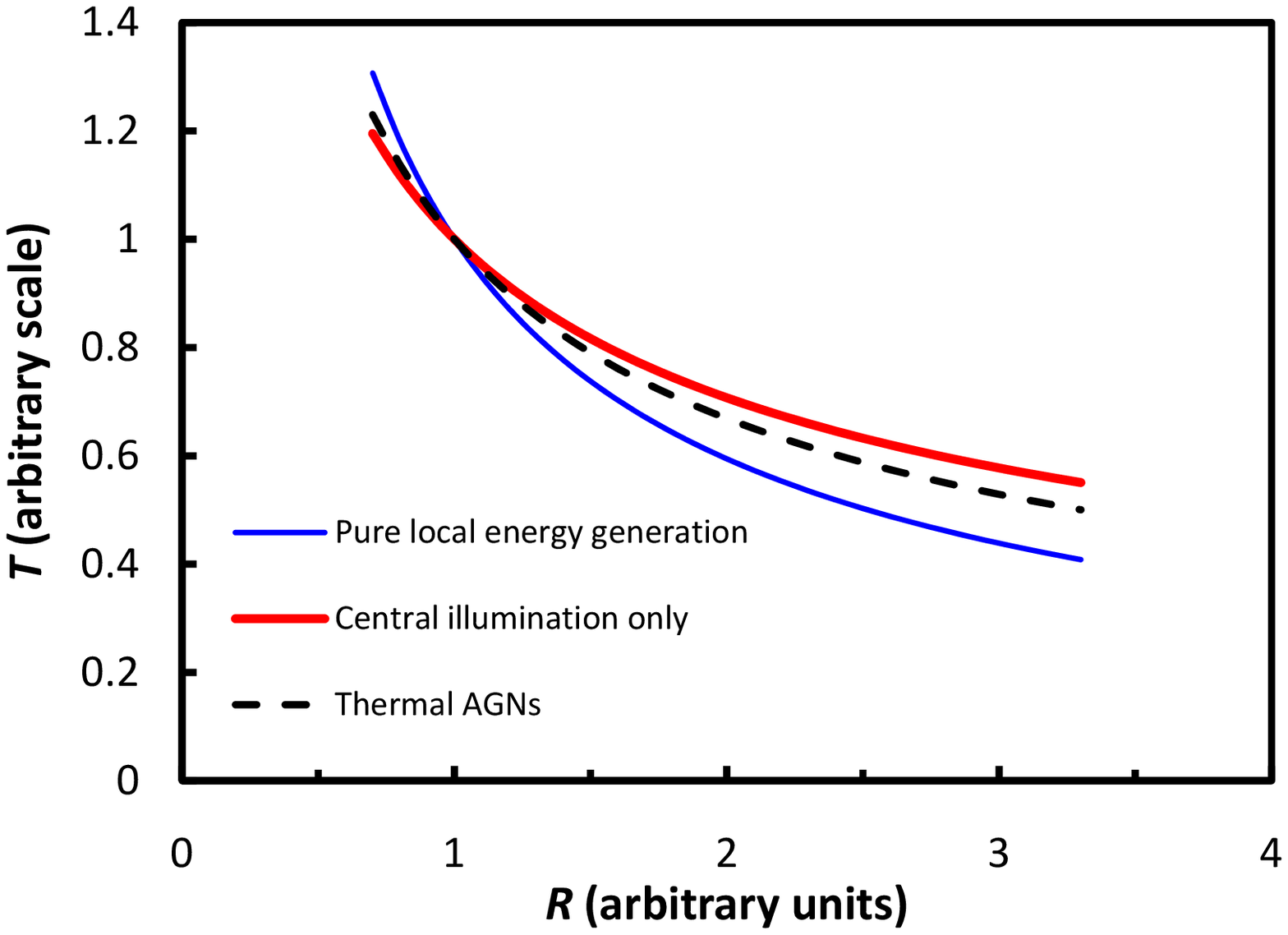,width=60mm,angle=0,clip=}
    \vspace{2mm}
  \end{minipage}
  \begin{minipage}[b]{65 mm}
    \psfig{figure=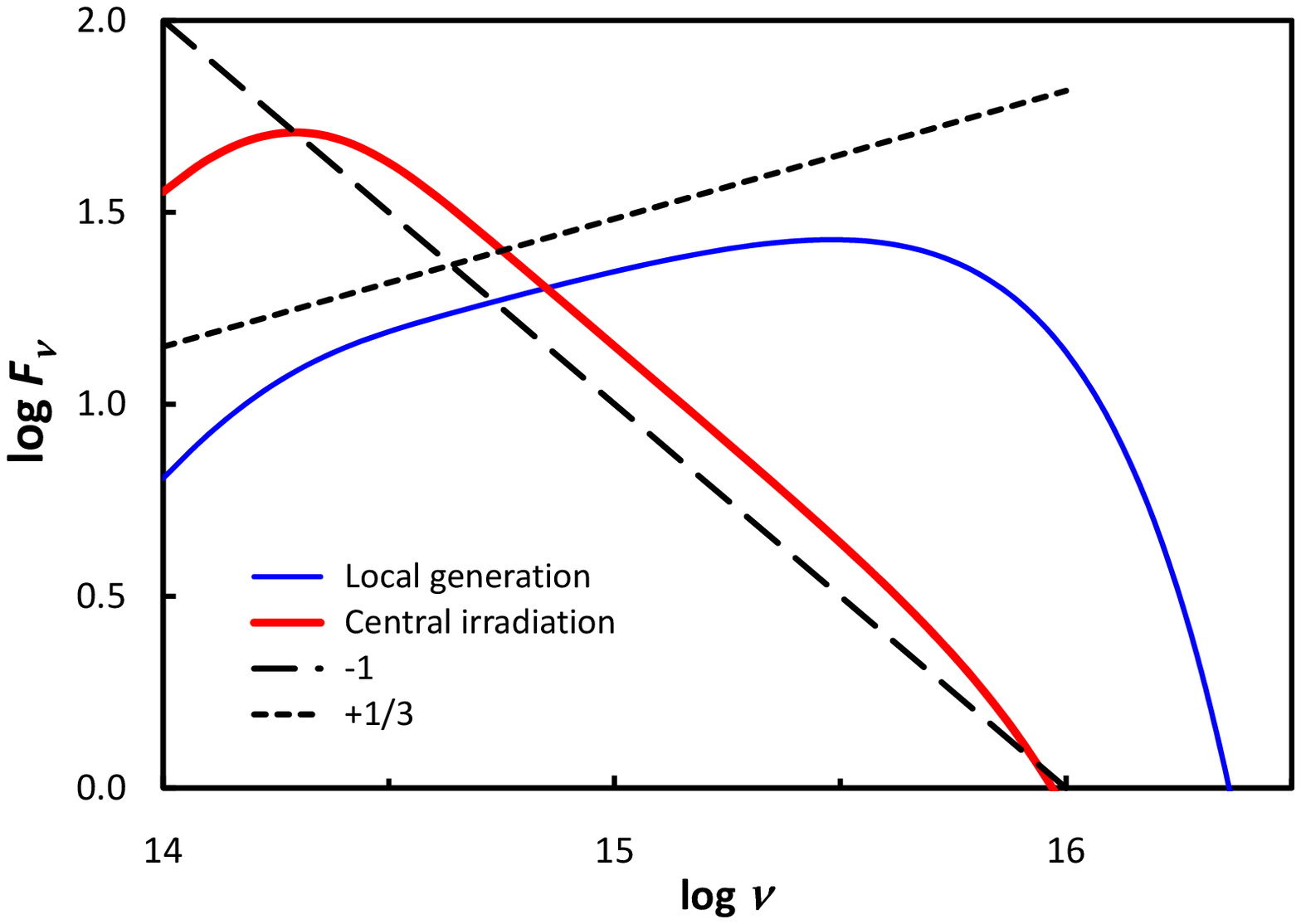,width=60mm,angle=0,clip=}
    \vspace{2mm}
  \end{minipage}
  \captionb{4}{\footnotesize {($a$)}~The temperature gradients in an accretion flow corresponding to purely
local energy generation (the case considered by Pringle \& Rees 1972 and Shakura \& Sunyaev 1973) and to central irradiation only.  The dashed line shows the temperature structure deduced from the observed spectral energy distribution of an AGN (Gaskell 2008 for details). {($b$)}~The spectra produced by truncated discs with the theoretical temperature structures shown in the left panel.  The dotted line shows the classic $F_{\nu} \propto \nu^{+1/3}$ accretion disc spectrum of PRSS for reference and the dashed line shows a $F_{\nu} \propto \nu^{-1}$ power-law.}
\end{figure}

The spectra in Fig.~4 were calculated by assuming that the gas at each radius radiates as a black body and adding up the relative contributions.  This is illustrated in left panel of Fig.~5. For purposes of illustration the smooth temperature gradient of the dashed curve in the left panel of Fig.~4 has been approximated as a series of steps in temperature.  The inner radius has been taken to give a cutoff in the spectral energy distribution around 4 Rydbergs (Mathews \& Ferland 1987).  Each curve corresponds to approximately a factor of two in radius.  The resultant integrated spectrum of the accretion flow is shown by the dashed line.  Notice that, at a given wavelength, the flux predominantly arises {\em from a given radius}.  For example, if we consider the flux at $\lambda$3000 ($\sim 10^{15}$ Hz) it can be seen that there are negligible contributions from the radius producing the extreme UV or the soft X-rays and from the radius producing the near IR.


\begin{figure}[htbp]
  \begin{minipage}[b]{65 mm}
  \psfig{figure=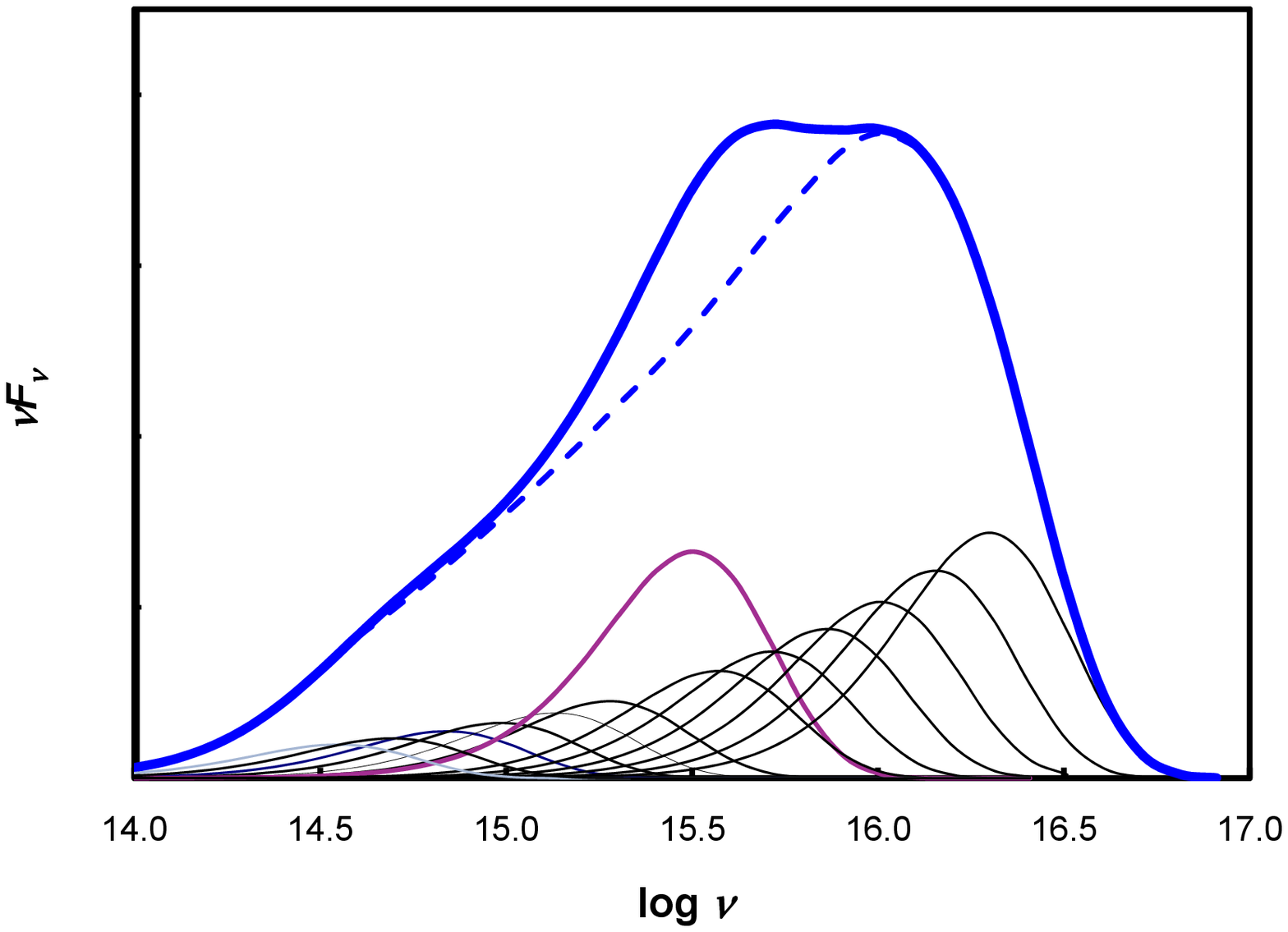,width=85mm,angle=0,clip=}
  \end{minipage}
  \begin{minipage}[b]{15 mm}
  \psfig{figure=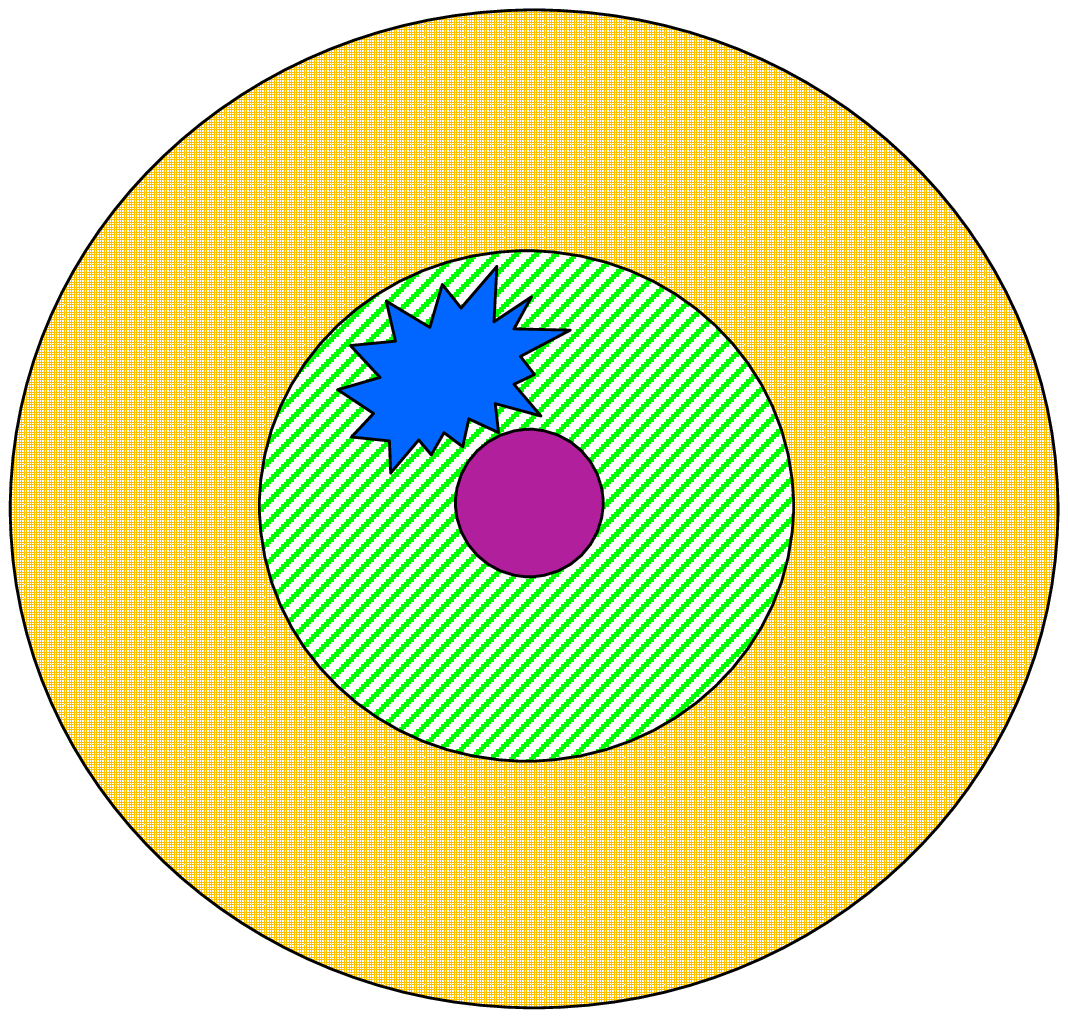,scale=0.33}
\vspace{13mm}
  \end{minipage}
\vspace{0.5mm}
\captionb{5}
{\footnotesize {\it (a)} The theoretical spectrum arising from an accretion flow.   The individual Planck curves (see text) are shown to scale. The dashed line is the integrated spectrum from a uniform disc.  The thick line shows the spectrum when the temperature of the zone producing the flux around 1 Rydberg (the thicker Planck curve) increases by 20\%.  {\it (b)} Cartoon of variability within an annulus.}
\label{}
\vspace{-0.4cm}
\end{figure}

It has long been known that AGNs are highly variable in the optical region of the spectrum and at higher energies (see, for example, Fig.~4 of Gaskell 2008).  Because the thermal emission at a particular wavelength comes from a limited range of radius, the variability at that wavelength must also be arising within a limited radius.  Fig.~5 illustrates how a flare at some radius only influences a limited spectral region.  We thus expect that the further apart spectral regions are in wavelength, the more independent is their variability.  This is indeed what is observed (see Gaskell 2008).  The independence of variability of different spectra regions (see Gaskell 2008) is important because it rules out variability at lower energies simply being the result of reprocessing of variability at higher energies.  Instead, there are clear cases of variability arising at lower energies without associated high energy variability (see, for example, Fig.~7 of Gaskell 2008).  Such variability must arise in an annulus, such as the central annulus of the cartoon shown in the right panel of Fig.~5.  The important point is that when variability is limited to an annulus, {\em the variability cannot be simultaneously around the annulus}.  There simply is not enough time for the annulus to coordinate its variability! Instead, as proposed in Gaskell (2008), the variability {\em must} be non-axisymmetric.

\sectionb{5}{CONSEQUENCES OF OFF-AXIS VARIABILITY}

Off-axis variability offers immediate solutions to all the problems enumerated in section 3.  Space here only permits a listing of the solutions with brief comments.  For more details the reader is referred to Gaskell (2010).

\begin{enumerate}
\item Off-axis emission naturally explains displaced broad-line peaks.  As an illustration of this, Fig.~6 shows how the most extreme asymmetric profile currently known (SDSS J0946+0139 -- Boroson \& Lauer 2010) can be readily fit with off-axis emission.  The active region responsible for the displaced red peak is on the receding side of the disc near the inner edge of the region producing H$\beta$.  The structure and kinematics are similar to the model of Fig.~2.  If the illumination is symmetric, the BLR shows the same profiles as in Fig.~2.  The off-axis model also easily explains changes in line profiles as the active areas orbit and turn off and on (see Gaskell 2010).


\begin{figure}[t!]
\vbox{
\centerline{\psfig{figure=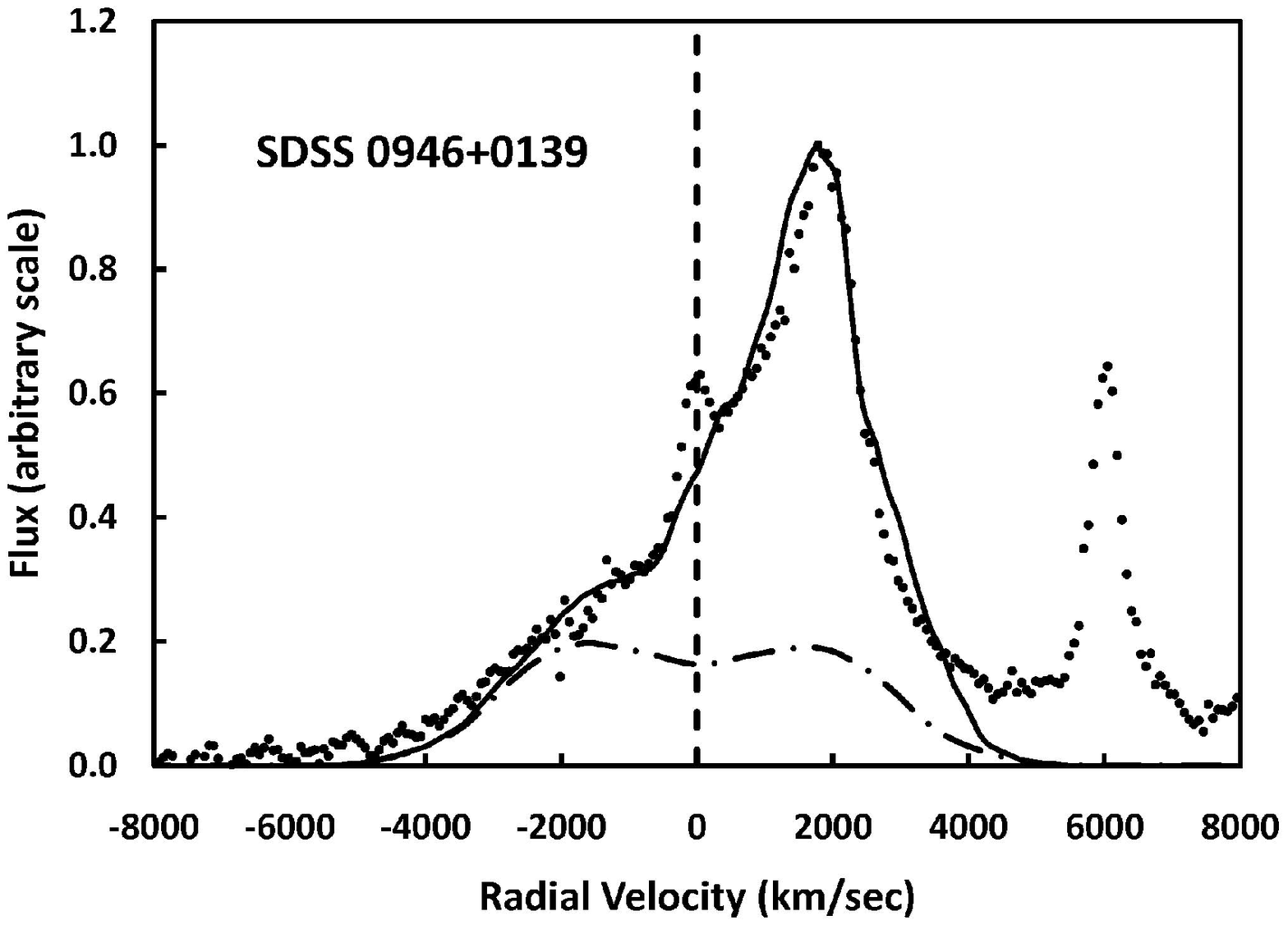,width=90mm,angle=0,clip=}}
\vspace{1mm}
\captionb{6}{\footnotesize The observed SDSS H$\beta$ profile of SDSS J0946+0139 (dots) compared with the off-axis illumination model (thick solid line) in which the GKN BLR has been illuminated from an active region located at the inner edge of the H$\beta$-emitting region. The NLR contribution to H$\beta$ to the observed spectum has not been subtracted.  The narrow emission line at +6000 km s$^{-1}$ is [O\,III] $\lambda$ 4959.  The dot-dashed line at the bottom is a similar off-axis profile to that shown in Fig.~1 (i.e., with central illumination and viewed from 30$^\circ$ off axis).  Figure from Gaskell (2010) where more details of the modeling can be found.}}
\label{}
\end{figure}

\item As has already been discussed in secion 5, off-axis variability explains why different spectral regions often vary independently.
\item Different continuum variability events giving different time delays are explained by differing locations of continuum events.  Those on the far side of the black hole will produce shorter delays, and those on the near side will produce longer delays.  This is quantitatively modeled in Gaskell (2010).
\item Off-axis variability explains how different kinematic signatures can be seen in velocity-resolved reverberation mapping and how different events can give different kinematic signatures in the same object.  A more rapid response of the blue side of a line simply means that the region of continuum variability is on the approaching side of the disc, rather than that the BLR is outflowing.
\item Off-axis variability produces strong variations in the parts of the disc closest to it. Because different regions of the disc produce different radial velocities, this explains how the most variable part of a line profile can have a surprisingly narrow range of velocity.
\item Similarly, localized variability can be independent of the total flux variability.  This explains the weak correlations or anti-correlations seen in narrow velocity regions in Fig.~3.
\item Finally, off-axis emission naturally produces velocity-dependent polarization.  Modelling shows (see Gaskell 2010) that the object-to-object differences in the changes in polarization across emission lines can naturally be explained by differing positions of the most active regions.  There is no need to invoke different kinematics or geometries in different AGNs.
\end{enumerate}

\sectionb{6}{DISCUSSION AND CONCLUSIONS}

Although other causes have been proposed for many of the phenomena enumerated in section 3, off-axis variability has the advantage of Occam's razor: one does not need to postulate different conditions in different objects. Instead, off-axis variability offers a simple common explanation of the phenomena.  It also provides support for our basic picture of AGNs and for the fundamental similarity of AGNs.
It does away with the need for exotic explanations, such as supermassive binary black holes Gaskell (1983).

The off-axis variability paradigm makes specific predictions about the details of variability that can be tested (see Gaskell 2010 for details).  The strongest prediction is that, since polarization is highly sensitive to departures from symmetry about the line of sight, the velocity dependence of polarization will change with time as different regions become active.

Obviously off-axis variability makes some things more complicated: it severely limits what we can learned from reverberation mapping and it makes looking for close binary black holes difficult since the profiles of a single BLR are complicated.  On the other hand, in the new paradigm, things such as changes in the velocity dependence of line polarization or the velocity dependence of the correlation of line and continuum flux variability (Fig.~3) enable us to learn about the distribution and frequency of occurrence of flares close to the BLR.

\thanks{I am grateful to the referee for useful comments and to the organizers of the 8th Serbian Conference on Spectral Line Shapes for the invitation to speak and for their hospitality.  This research has been supported by US National Science Foundation grants AST 03-07912 and AST 08-03883, NASA grant NNH-08CC03C, grant 32070017 of the GEMINI-CONICYT Fund.}

\References

\refb Antonucci, R.\ 1993, ARA\&A, 31, 473

\refb Antonucci, R.~R.~J. 2011, Astron. \& Astrophys. Transactions, in press.

\refb Axon, D.~J., Robinson, A., Young, S., Smith, J.~E., \& Hough, J.~H.\ 2008, Mem. Soc. Astron. Italiana, 79, 1213

\refb Blandford, R.~D. \& Rees, M.~J. 1978, in {\it Pittsburgh Conference on BL Lac Objects} (Pittsburgh: University of Pittsburgh), p. 328

\refb Boroson, T.~A., \& Lauer, T.~R.\ 2010, AJ, 140, 390

\refb Clavel, J., Wamsteker, W. \& Glass, I.~S. 1989, ApJ, 337, 236

\refb Curtis, H. D. 1918, Lick Observatory Pubs., Vol. 13, p.9

\refb Denney, K.~D., et al.\ 2009b, ApJ. Letts. 704, L80

\refb Eracleous, M. \& Halpern, J.~P. 2003, ApJ., 599, 886

\refb Gaskell, C.~M.\ 1983, Liege International Astrophys. Colloq., 24, 473.

\refb Gaskell, C.~M. 1996, in {\it Jets from Stars and Galactic Nuclei},
ed. W. Kundt. (Berlin: Springer-Verlag), p. 165

\refb Gaskell, C.~M.\ 2006, Astron. Soc. Pacific Conf. Ser., 360, 111

\refb Gaskell, C.~M.\ 2008, Rev. Mexicana Astron. Ap. Conf. Ser., 32, 1

\refb Gaskell, C.~M. 2009, New Astron. Rev, 53, 140.

\refb Gaskell, C.~M. \& Goosmann, R.~W.\ 2008, ApJ,~submitted [arXiv:0805.4258]

\refb Gaskell, C.~M., Goosmann, R.~W., \& Klimek, E.~S.\ 2008, Mem. Soc. Astron. Italiana, 79, 1090

\refb Gaskell, C.~M., Klimek, E.~S., \& Nazarova, L.~S.\ 2007, ApJ, submitted (GKN) [arXiv:0711.1025]

\refb Jennison, R.~C. \& Das Gupta, M.~K. (1953) Nature, 172, 996.

\refb Keel, W.~C. 1980, AJ, 85, 198.

\refb Kollatschny, W. \& Dietrich, M.\ 1996, A\&A, 314, 43.

\refb Lynden-Bell 1969, Nature, 223, 690.

\refb Maoz, D.\ 1994, Astron. Soc. Pacific Conf. Ser. 69, 95

\refb Mathews, W.~G. \& Ferland, G.~J. 1987, ApJ, 323, 456.

\refb Osterbrock, D.~E.\ 1978, Proc. Nat. Acad. Sci., 75, 540.

\refb Pringle, J.~E., \& Rees, M.~J.\ 1972, A\&A, 21, 1

\refb Sergeev, S.~G., Pronik, V.~I., \& Sergeeva, E.~A.\ 2001, ApJ, 554, 245

\refb Shakura, N.~I., \& Syunyaev, R.~A.\ 1973, A\&A, 24, 337.

\refb Suganuma, M., et al.\ 2006, ApJ, 639, 46


\end{document}